\RequirePackage{fixltx2e}
\RequirePackage{fix-cm}

\documentclass[%
floatfix,
showkeys,
twocolumn, %
nofootinbib, %
superscriptaddress, %
]{revtex4-1}

\usepackage{cmap}

\usepackage{ucs}
\usepackage[utf8x]{inputenc}
\usepackage[T1,T2A]{fontenc}
\usepackage[german,russian,english]{babel}

\usepackage[sort&compress]{natbib}
\usepackage{amsmath}
\usepackage{amssymb}

\usepackage{mathtools}
\mathtoolsset{
showonlyrefs,
mathic = true
}

\allowdisplaybreaks

\usepackage{hyperref}
\hypersetup{backref,
 colorlinks=false}
\hypersetup{pdfborder=0 0 0}

\usepackage{microtype}
\UseMicrotypeSet[protrusion]{alltext}

\usepackage{graphicx}

\usepackage[scanall]{psfrag}

\usepackage{listings}
\usepackage{listingsutf8}
\usepackage{fixltx2e}

\usepackage{nicefrac}

\makeatletter
\def\ps@pprintTitle{%
     \let\@oddhead\@empty
     \let\@evenhead\@empty
     \let\@oddfoot\@empty
     \let\@evenfoot\@oddfoot}
\makeatother

\lstnewenvironment{python}[1][]{\lstset{language=python,
   columns=fullflexible,
   xleftmargin=1em,
   belowskip=0pt,
   tabsize=4,
   breaklines=true,
   basicstyle=\ttfamily,
   comment=[f][0]{\#},
   literate=::1,
   prebreak=\space,
    postbreak=\raisebox{0ex}[0ex][0ex]{\ensuremath{\hookrightarrow\space}},
   breakautoindent=true,
   breakatwhitespace=false,
   keywordstyle={}
}}{}

\usepackage{changepage}

\newcommand\brwrap[3]{%
  \setbox0=\hbox{$#2$}
  \left#1\vbox to \the\ht0{\hbox to 0pt{}}\right.\kern-.2em
  \begingroup #2\endgroup\kern-.15em
  \left.\vbox to \the\ht0{\hbox to 0pt{}}\right#3
}

\usepackage{physics}
\DeclareMathOperator{\diag}{diag}

\begin{document}
\graphicspath{{image/cadabra2/en/}{image/cadabra2/}}

\title{New Features in the Second Version of the Cadabra Computer Algebra System}

\author{D. S. Kulyabov}
\email{kulyabov-ds@rudn.ru}
\affiliation{Department of Applied Probability and Informatics,\\
  Peoples' Friendship University of Russia (RUDN University),\\
  6 Miklukho-Maklaya St, Moscow, 117198, Russian Federation}
\affiliation{Laboratory of Information Technologies\\
  Joint Institute for Nuclear Research\\
  6 Joliot-Curie, Dubna, Moscow region, 141980, Russia}

\author{A. V. Korolkova}
\email{korolkova-av@rudn.ru}
\affiliation{Department of Applied Probability and Informatics,\\
  Peoples' Friendship University of Russia (RUDN University),\\
  6 Miklukho-Maklaya St, Moscow, 117198, Russian Federation}

\author{L. A. Sevastianov}
\email{sevastianov-la@rudn.ru}
\affiliation{Department of Applied Probability and Informatics,\\
  Peoples' Friendship University of Russia (RUDN University),\\
  6 Miklukho-Maklaya St, Moscow, 117198, Russian Federation}
\affiliation{Bogoliubov Laboratory of Theoretical Physics\\
  Joint Institute for Nuclear Research\\
  6 Joliot-Curie, Dubna, Moscow region, 141980, Russia}

\begin{abstract}
In certain scientific domains, there is a need for tensor
operations. To facilitate tensor computations, computer algebra
systems are employed. In our research, we have been using Cadabra as
the main computer algebra system for several years. Recently, an
operable second version of this software was released.
In this version, a number of improvements were made that can be regarded as revolutionary ones. The most
significant improvements are the implementation of component computations and the change in the ideology
of the Cadabra’s software mechanism as compared to the first version. This paper provides a brief overview
of the key improvements in the Cadabra system.
\end{abstract}


\maketitle

\section{Introduction}
\label{sec:intro}

  Computer algebra systems that support tensor calculus can be
  conventionally divided into three
  groups~\cite{maccallum:2018:computer-algebra-gravity}. The first
  group includes universal tensor calculus systems designed mainly for
  use in the general theory of relativity. In these systems, focus is
  placed on computing characteristic quantities of Riemannian geometry
  (Christoffel symbols, Riemann
  tensor)~\cite{ilyin:1996:atensor,gomez-lobo:2012:spinors-xact,maccallum:2002:ca-gr}.
  The second group includes computer algebra systems for tensor
  computations in quantum field
  theory~\cite{bolotin:2013:redberry,poslavsky:2015:redberry,fliegner:1999:form,heck:2000:formbook,tung:2005:form}.
  They focus on computations with Dirac spinors and simple
  symmetries. The third group of computer algebra systems for tensor
  calculus is designed mainly as a framework for arbitrary tensor
  computations. These systems have fewer default templates but provide
  more expressive means for designing new objects. It is to this group
  that Cadabra belongs. This paper is devoted to the specific features
  of this computer algebra system.

  Cadabra is currently available in two versions. The first version
  (hereinafter, Cadabra 1.x) significantly differs from the other
  computer algebra systems for tensor calculus, particularly, in
\begin{itemize}
\item  use of the TEX notation;
\item  convenient definition of tensors of arbitrary types;
\item  use of Young tableau to describe the symmetry
properties of tensors.
\end{itemize}

However, Cadabra 1.x has significant disadvantages. In particular,
this version does not support component computations. Moreover, the
monolithic and rigid software structure of the system held no promise
for the implementation of component computations in the near
future. However, a revolutionary step taken in the second version of
the system (hereinafter, Cadabra 2.x), namely, the use of the Python
ecosystem, made it possible to solve these problems. In our opinion,
it is the combination of component computations and Python ecosystem
that constitutes a revolutionary improvement of the Cadabra system.

Unfortunately, the documentation on Cadabra leaves much to be
desired. The website of the system (\url{https://cadabra.science}) provides
only a few examples. More specific information can be found in
research papers that may contain possible applications of
Cadabra~\cite{peeters:2007:cadabra,peeters:2007:cpc:cadabra,brewin:2010:cpc:tensor-cadabra,kulyabov:2009:nucleilett:cadabra,kulyabov:2013:springer:cadabra,kulyabov:2016:pcs}.

  This paper provides a brief overview of the new features in Cadabra
  2.x. The paper is organized as follows.
  Section~\ref{sec:realisation} discusses the implementation of
  Cadabra 1.x and 2.x. Section~\ref{sec:syntax} describes the process
  of action variation for a source-free electromagnetic field to
  illustrate the syntax of Cadabra 2.x. To compare the 1.x and 2.x
  syntaxes,
  see~\cite{kulyabov:2009:nucleilett:cadabra,kulyabov:2013:springer:cadabra,kulyabov:2016:pcs}. Section~\ref{sec:sympy}
  considers one of the most important features in Cadabra 2.x: its
  transparent interaction with the universal scalar computer algebra
  system SymPy. Section~\ref{sec:coord} describes the key (in our
  opinion) innovation in Cadabra 2.x---component computations---by the
  example of finding basic quantities for the general theory of
  relativity, namely, Christoffel connection and different
  curvatures. Finally, Section~\ref{sec:plot} illustrates work with
  graphics in Cadabra 2.x.

  When implementing the examples, we used code fragment from the
  Cadabra 2.x documentation
  (\url{https://cadabra.science/tutorials.html}).

  \section{Implementation Features of Cadabra 1.x and 2.x}

\label{sec:realisation}

Each computer algebra system has its own implementation
features. There are several levels of implementation:
\begin{itemize}
\item  notation;
\item  manipulation language;
\item  implementation language;
\item  extension language.
\end{itemize}

Not every computer algebra system has all these
levels. For instance, in most systems, the notation is
based on the manipulation language.  

In Cadabra 2.x (as in 1.x), the notation is based on
the \TeX{} notation (more precisely, \TeX{}-like notation).
In this case, a certain subset of \TeX{} symbols (letters of
different alphabets, symbols for integral and derivatives, etc.) is used. Above all, indices are denoted by
the symbols \verb|_| and \verb|^|, as in the \TeX{} system.

The manipulation language is used to work in the system. The syntax of
Cadabra 1.x is quite simple and more oriented to code parsing rather
than to userfriendliness. This approach to language design was common
in the early years of computing technology development (e.g., Shell
and Perl languages). Cadabra 2.x takes a qualitative step forward by
switching to Python. As a result, all operations in Cadabra 2.x are
written in a Python-like syntax.

Cadabra 1.x is implemented in C++ with the use of
the LiE computer algebra system~\cite{leeuwen:1992:lie} (currently, the
compilation of the LiE system causes certain difficulties). Its main purpose is processing Lie groups, on
which operations with tensor symmetries are based.
It should be noted that the entire system was actually
implemented by one person. This required great
efforts. However, the presence of only one author and
the monolithic software structure of the system caused
some concern.

In Cadabra 2.x, the author fundamentally changed
his approach to the structure of the system by integrating it with the Python ecosystem. The 2.x system is still
written in C++, but Python is used as a glue language
(and also as a manipulation language). In addition,
Python can be employed for writing extensions.

The Python infrastructure opens access to a large
number of scientific libraries, including the SciPy
project~\cite{oliphant:2007:scipy}. This allows the SciPy libraries to be used
seamlessly, transparently to the user. In our opinion,
this allowed Cadabra 2.x to take a revolutionary step
forward.

\section{Elements of the 2.x Syntax}
\label{sec:syntax}

Let us recall some elements of the 1.x and 2.x syntaxes. As an
example, we obtain the source-free Maxwell's equation~\cite{ll:2::en}:
\begin{equation}
  \label{eq:F_mn}
  \partial_{\mu} F^{\mu \nu} = 0,
\end{equation}
by varying the action
\begin{equation}
  \label{eq:S-maxwell}
  S = - \frac{1}{4} \int F_{\mu\nu} F^{\mu\nu} \dd{x}.
\end{equation}

In this case, the Maxwell tensor $F_{\mu \nu}$ is expressed in
terms of the vector potential:
\begin{equation}
  \label{eq:F=A-A}
  F_{\mu\nu} = \partial_{\mu}{A_{\nu}} - \partial_{\nu}{A_{\mu}}.
\end{equation}

Again, both versions of Cadabra use the \TeX{}
notation. In Cadabra 2.x, it becomes possible to use functionality of Python, in particular, define functions
directly in the text of a program written in Cadabra 2.x.
Note that, just like Cadabra 1.x, the 2.x version does
not process resulting expressions by default (except for
collecting like terms); this processing is left to the user.
If it is required to apply a certain set of rules to all
expressions used, then the function \verb|post_process|
can be called, which is executed after each operation
(in fact, \verb|post_process| is simply a Python function):
\begin{python}
def post_process(ex):
    sort_product(ex)
    canonicalise(ex)
    collect_terms(ex)
\end{python}

If necessary, this function can be made empty, thus
disabling any processing of expressions.
The interface of Cadabra 2.x exploits the ideology
of a notepad, i.e., in addition to writing program code,
the user can also add comments. However, in contrast
to iPython~\cite{perez:2007:ipython}, text is written in the~\LaTeX{} syntax
rather than in the Markdown syntax~\cite{rfc:7763}.

The object in Cadabra 2.x can be assigned a property, which, in turn, has a set of its own settings. Since
we are dealing with tensors, the most useful property is
\verb|Indices|. The option \verb|position=free| allows the
system to raise and lower indices:
\begin{python}
{\mu,\nu,\rho}::Indices(position=free).
x::Coordinate.
\partial{#}::Derivative.
\end{python}

Here, \verb|#| is a wildcard. The dot at the end of the
expression suppresses the output (as is common in
computer algebra systems).

To work with abstract indices, it is necessary to take
into account the symmetry properties of tensors.
In addition, when differentiating and integrating, the
coordinate dependence of objects needs to be taken
into account:
\begin{python}
F_{\mu\nu}::AntiSymmetric;
F_{\mu\nu}::Depends(x).
A_{\mu}::Depends(x,\partial{#}).
\delta{#}::Accent;
\end{python}
\begin{adjustwidth}{1em}{0cm}${}\text{Attached property AntiSymmetric to~}F_{\mu \nu}.$\end{adjustwidth}
\begin{adjustwidth}{1em}{0cm}${}\text{Attached property Accent to~}\delta{\#}.$\end{adjustwidth}

In this case, the variation sign $\delta$ is regarded as a
modifier (rather than as an object with the \verb|Derivative| property) and does not introduce any additional
computational semantics.
In our example, $F_{\mu\nu}$ is a Maxwell tensor. We express
it in terms of the vector potential $A_{\mu}$~\eqref{eq:F=A-A}:
\begin{python}
F:= F_{\mu\nu} = \partial_{\mu}{A_{\nu}} - \partial_{\nu}{A_{\mu}};
\end{python}
\begin{adjustwidth}{1em}{0cm}${}F_{\mu \nu} = \partial_{\mu}{A_{\nu}}-\partial_{\nu}{A_{\mu}}$\end{adjustwidth}

Here, the sign $:=$ defines a row label (in our case, it
is $F$). The label denominates the expression for convenience of
referring to it.

Next, the action for the electromagnetic field~\eqref{eq:S-maxwell} is
defined:  
\begin{python}
S:= -1/4 \int{ F_{\mu\nu} F^{\mu\nu} }{x};
\end{python}
\begin{adjustwidth}{1em}{0cm}${} - \frac{1}{4}\int F^{\mu \nu} F_{\mu \nu}\,\,{\rm d}x$\end{adjustwidth}

By substitution, the action is expressed in terms of
the vector potential $A_{\mu}$:
\begin{python}
substitute(S, F);
\end{python}
\begin{adjustwidth}{1em}{0cm}${} - \frac{1}{4}\int \brwrap{(}{\partial_{\mu}{A_{\nu}}-\partial_{\nu}{A_{\mu}}}{)} \brwrap{(}{\partial^{\mu}{A^{\nu}}-\partial^{\nu}{A^{\mu}}}{)}\,\,{\rm d}x$\end{adjustwidth}
Then, the action has the following form:  
\begin{python}
S;
\end{python}
\begin{adjustwidth}{1em}{0cm}${} - \frac{1}{4}\int \brwrap{(}{\partial_{\mu}{A_{\nu}}-\partial_{\nu}{A_{\mu}}}{)} \brwrap{(}{\partial^{\mu}{A^{\nu}}-\partial^{\nu}{A^{\mu}}}{)}\,\,{\rm d}x$\end{adjustwidth}

The value of the action in this expression differs
from the initial one. It takes the form that it received
after the last computation. This is a bit unusual. The
point is that most computer algebra systems are implemented using functional languages, or they follow a
functional paradigm in which variables have the property of immutability. In this case, the label acts as a
variable in imperative languages (Python is an imperative
language). This makes work in Cadabra 2.x necessary linear: the user
cannot randomly navigate
through the notebook and perform computations at
arbitrary points.

Let us vary the action:  
\begin{python}
vary(S, $A_{\mu} -> \delta{A_{\mu}}$);
\end{python}
\begin{adjustwidth}{1em}{0cm}${} - \frac{1}{4}\int \brwrap{(}{\brwrap{(}{\partial^{\mu}{A^{\nu}}-\partial^{\nu}{A^{\mu}}}{)} \brwrap{(}{\partial_{\mu}{\delta{A_{\nu}}}-\partial_{\nu}{\delta{A_{\mu}}}}{)}+\brwrap{(}{\partial_{\mu}{A_{\nu}}-\partial_{\nu}{A_{\mu}}}{)} \brwrap{(}{\partial^{\mu}{\delta{A^{\nu}}}-\partial^{\nu}{\delta{A^{\mu}}}}{)}}{)}\,\,{\rm d}x$\end{adjustwidth}

Here, the expression itself, rather than its label, is
used. For this purpose, the expression is put between
two dollar symbols (\$), as in the standard \TeX{}.

Next, we expand the products and collect the like
terms:
\begin{python}
distribute(S);
\end{python}
\begin{adjustwidth}{1em}{0cm}${} - \frac{1}{4}\int \brwrap{(}{4\partial^{\mu}{A^{\nu}} \partial_{\mu}{\delta{A_{\nu}}}-4\partial^{\mu}{A^{\nu}} \partial_{\nu}{\delta{A_{\mu}}}}{)}\,\,{\rm d}x$\end{adjustwidth}

Then, integration by parts is carried out:
\begin{python}
integrate_by_parts(S, $\delta{A_{\mu}}$);
\end{python}
\begin{adjustwidth}{1em}{0cm}${} - \frac{1}{4}\int \brwrap{(}{-4\delta{A^{\mu}} \partial^{\nu}\brwrap{(}{\partial_{\nu}{A_{\mu}}}{)}+4\delta{A^{\mu}} \partial^{\nu}\brwrap{(}{\partial_{\mu}{A_{\nu}}}{)}}{)}\,\,{\rm d}x$\end{adjustwidth}

In this case, integration by parts is quite a formal
action that uses only the \verb|Derivative| property of
the object.

Once again, we perform the substitution, expand
the products, and collect the like terms:  
\begin{python}
substitute(_, $\partial_{\mu}{A_{\nu}} -> 1/2 \partial_{\mu}{A_{\nu}} + 1/2 F_{\mu\nu} + 1/2 \partial_{\nu}{ A_{\mu}}$);
\end{python}
\begin{adjustwidth}{1em}{0cm}${} - \frac{1}{4}\int \brwrap{(}{-4\delta{A^{\mu}} \partial^{\nu}\brwrap{(}{\frac{1}{2}\partial_{\nu}{A_{\mu}} - \frac{1}{2}F_{\mu \nu}+\frac{1}{2}\partial_{\mu}{A_{\nu}}}{)}+4\delta{A^{\mu}} \partial^{\nu}\brwrap{(}{\frac{1}{2}\partial_{\mu}{A_{\nu}}+\frac{1}{2}F_{\mu \nu}+\frac{1}{2}\partial_{\nu}{A_{\mu}}}{)}}{)}\,\,{\rm d}x$\end{adjustwidth}
\begin{python}
distribute(_);
\end{python}
\begin{adjustwidth}{1em}{0cm}${}-\int \delta{A^{\mu}}
  \partial^{\nu}{F_{\mu \nu}}\,\,{\rm d}x$\end{adjustwidth}

Here, the label \verb|_| denotes the previous expression.
  
As a result, we obtain the desired Maxwell equation~\eqref{eq:F_mn}:
\begin{equation}
  \partial^\nu F_{\mu\nu}=0
\end{equation}

This example demonstrates that the syntax of the
manipulation language in Cadabra 2.x is based on the
Python syntax, which is more customary than the syntax of the 1.x language.

\section{Interaction between Cadabra and SymPy}
\label{sec:sympy}

  Computer algebra systems for tensor calculus support quite a small
  number of operations. They are sufficient for basic manipulations
  with tensors in the formalism of abstract indices, as well as the
  index-free formalism. However, in many cases (e.g., full-fledged
  implementation of component computations), the support of scalar
  operations is required. If a computer algebra system for tensor
  calculus is implemented in the framework of a universal computer
  algebra system, then no problems arise. However, Cadabra is an
  independent system. Cadabra 1.x implements a mechanism (though
  inconvenient) for communication with the Maxima universal computer
  algebra system; however, it seems that this mechanism was
  implemented only as a proof of concept.

In Cadabra 2.x, communication with the universal
computer algebra system is implemented via
SymPy~\cite{lamy:sympy_starter}.
Moreover, this communication is seamless: the
work of the mechanism is invisible for the user, which
is owing to implementation of Cadabra 2.x in Python.

Let us illustrate the use of SymPy in Cadabra 2.x by
computing the integral
\begin{equation}
  \label{eq:int-sympy}
  \int \frac{1}{x} \dd{x}.
\end{equation}

The main function for the explicit call of SymPy is
\verb|map_sympy()|. This function has a side effect: it
changes the value of the argument. However, as noted
above, the absence of immutability is a feature of
Cadabra 2.x. Let us consider the simplest call of this
function:
\begin{python}
ex := \int{1/x}{x};
\end{python}
\begin{adjustwidth}{1em}{0cm}${}\int {x}^{-1}\,\,{\rm d}x$\end{adjustwidth}
\begin{python}
map_sympy(_);
\end{python}
\begin{adjustwidth}{1em}{0cm}${}\log\brwrap{(}{x}{)}$\end{adjustwidth}

To confirm the presence of this side effect, we
check the current value of the expression \verb|ex|:
\begin{python}
ex;
\end{python}
\begin{adjustwidth}{1em}{0cm}${}\log\brwrap{(}{x}{)}$\end{adjustwidth}

It can be seen that the value of \verb|ex| has changed.

We can transfer the value of the expression to
SymPy and, moreover, call a particular function to
process it. For instance, in this case, we can call the
SymPy's function \verb|integrate|:
\begin{python}
ex := 1/x;
\end{python}
\begin{adjustwidth}{1em}{0cm}${}{x}^{-1}$\end{adjustwidth}

The second argument of the function \verb|map_sympy()| is SymPy
function name:
\begin{python}
map_sympy(ex,"integrate");
\end{python}
\begin{adjustwidth}{1em}{0cm}${}\log\brwrap{(}{x}{)}$\end{adjustwidth}
\begin{python}
ex;
\end{python}
\begin{adjustwidth}{1em}{0cm}${}\log\brwrap{(}{x}{)}$\end{adjustwidth}

Again, this confirms the
side effect of
\verb|map_sympy()|.

In Python, the same action can be performed in
several ways. Hence, this can be done in Cadabra 2.x.
For illustration purposes, let us consider the following
variants.

The class method \verb|_sympy_()| can be used as follows:
\begin{python}
ex := \int{1/x}{x};
\end{python}
\begin{adjustwidth}{1em}{0cm}${}\int {x}^{-1}\,\,{\rm d}x$\end{adjustwidth}

While regarding the label \verb|ex| as an object, we call
the method \verb|_sympy_()|:
\begin{python}
ex._sympy_();
\end{python}
\begin{adjustwidth}{1em}{0cm}${}\log{\left (x \right)}$\end{adjustwidth}

Check the state of the environment:
\begin{python}
ex;
\end{python}
\begin{adjustwidth}{1em}{0cm}${}\int {x}^{-1}\,\,{\rm d}x$\end{adjustwidth}

It is seen that the state of the environment has not
changed, i.e., the method \verb|_sympy_()| has no side
effect.

In addition, we can use the function \verb|sympy| with a
method corresponding to a callee function of the
SymPy environment:
\begin{python}
ex := 1/x;
\end{python}
\begin{adjustwidth}{1em}{0cm}${}{x}^{-1}$\end{adjustwidth}

Let us call the function \verb|sympy| with the method
\verb|integrate|:
\begin{python}
sympy.integrate(ex);
\end{python}
\begin{adjustwidth}{1em}{0cm}${}\log{\left (x
    \right)}$\end{adjustwidth}

Note that this function always requires specifying a
particular method, which is why it cannot be used in
the previous case.

Again, check the state of the environment:
\begin{python}
ex;
\end{python}
\begin{adjustwidth}{1em}{0cm}${}{x}^{-1}$\end{adjustwidth}

It can be seen that the function \verb|sympy| does not
have the side effect.
Based on the examples considered above, we can
conclude that the interaction with SymPy in Cadabra
2.x is implemented in quite an elegant way. In our
opinion, the main advantage of this operation is its
deep integration with the system, e.g., for implementation of component computations (see Section~\ref{sec:coord}).

  \section{Component Computations in Cadabra~2.x}
\label{sec:coord}

To illustrate component operations, we find curvature $R$ on a sphere $S^{2}$ of radius $r$:
\begin{equation}
  g_{\alpha\beta} = \diag\qty(r^2,r^2 \sin^2 \vartheta).
\end{equation}

For this purpose, we evaluate Christoffel symbols $\Gamma^{\alpha}_{\mu\nu}$,
Riemann tensor $R^{\alpha}\,_{\beta \mu \nu}$, and Ricci tensor $R_{\alpha\beta}$~\cite{mtw:1::en,ll:2::en}.
Let us define coordinates and labels for the indices
while specifying what values these labels can take:
\begin{python}
{\theta, \varphi}::Coordinate;
{\alpha, \beta, \gamma, \delta, \rho, \sigma, \mu, \nu, \lambda}::Indices(values={\varphi, \theta}, position=fixed);
\partial{#}::PartialDerivative;
\end{python}
\begin{adjustwidth}{1em}{0cm}${}\text{Attached property Coordinate to~}\brwrap{[}{\theta,~\discretionary{}{}{} \varphi}{]}.$\end{adjustwidth}
\begin{adjustwidth}{1em}{0cm}${}\text{Attached property Indices(position=fixed) to~}\brwrap{[}{\alpha,~\discretionary{}{}{} \beta,~\discretionary{}{}{} \gamma,~\discretionary{}{}{} \delta,~\discretionary{}{}{} \rho,~\discretionary{}{}{} \sigma,~\discretionary{}{}{} \mu,~\discretionary{}{}{} \nu,~\discretionary{}{}{} \lambda}{]}.$\end{adjustwidth}
\begin{adjustwidth}{1em}{0cm}${}\text{Attached property PartialDerivative to~}\partial{\#}.$\end{adjustwidth}

Next, we define a tensor $g$ with the \verb|Metric| property and, similarly, define the inverse metric:
\begin{python}
g_{\alpha\beta}::Metric.
g^{\alpha\beta}::InverseMetric.
\end{python}

In this case, it is sufficient to specify the components for the original metric. The components for the
inverse metric are computed using the function \verb|complete|:
\begin{python}
g:={ g_{\theta\theta} = r**2, g_{\varphi\varphi} = r**2 \sin(\theta)**2 }.
complete(g, $g^{\alpha\beta}$);
\end{python}
\begin{adjustwidth}{1em}{0cm}${}\brwrap{[}{g_{\theta \theta} = {r}^{2},~\discretionary{}{}{} g_{\varphi \varphi} = {r}^{2} {\brwrap{(}{\sin{\theta}}{)}}^{2},~\discretionary{}{}{} g^{\varphi \varphi} = {\brwrap{(}{{r}^{2} {\brwrap{(}{\sin{\theta}}{)}}^{2}}{)}}^{-1},~\discretionary{}{}{} g^{\theta \theta} = {r}^{-2}}{]}$\end{adjustwidth}

Let us define the Christoffel symbols $\Gamma^{\alpha}_{\mu\nu}$:
\begin{python}
Gamma:= \Gamma^{\alpha}_{\mu\nu} = 1/2 g^{\alpha\beta} (\partial_{\nu}{g_{\beta\mu}} + \partial_{\mu}{g_{\beta\nu}} - \partial_{\beta}{g_{\mu\nu}});
\end{python}
\begin{adjustwidth}{1em}{0cm}${}\Gamma^{\alpha}\,_{\mu \nu} = \frac{1}{2}g^{\alpha \beta} \brwrap{(}{\partial_{\nu}{g_{\beta \mu}}+\partial_{\mu}{g_{\beta \nu}}-\partial_{\beta}{g_{\mu \nu}}}{)}$\end{adjustwidth}

We expand the Christoffel symbols by using the
metric tensor. In this case, expansion is applied only to
the right-hand side of the definition (after the equal
sign). This order of the expansion is determined by the
option \verb|rhsonly=True|. The function \verb|evaluate()|
implicitly calls SymPy for operations on components
(this is another advantage of using the Python infrastructure). To expand trigonometric relationships,
SymPy should be called explicitly:
\begin{python}
evaluate(Gamma, g, rhsonly=True)
map_sympy(Gamma, "expand_trig");
\end{python}
\begin{adjustwidth}{1em}{0cm}${}\Gamma^{\alpha}\,_{\mu \nu} = \square{}_{\mu}{}_{\nu}{}^{\alpha}\brwrap{\{}{\begin{aligned}\square{}_{\varphi}{}_{\theta}{}^{\varphi}= & {\brwrap{(}{\tan{\theta}}{)}}^{-1}\\[-.5ex]
\square{}_{\theta}{}_{\varphi}{}^{\varphi}= & {\brwrap{(}{\tan{\theta}}{)}}^{-1}\\[-.5ex]
\square{}_{\varphi}{}_{\varphi}{}^{\theta}= & -\sin{\theta} \cos{\theta}\\[-.5ex]
\end{aligned}}{.}$
\end{adjustwidth}

Similarly, for the Riemann tensor $R^{\alpha}\,_{\beta \mu \nu}$, we write its
definition and evaluate its components:
\begin{python}
R4:= R^{\rho}_{\sigma\mu\nu} = +\partial_{\mu}{\Gamma^{\rho}_{\sigma\nu}} - \partial_{\nu}{\Gamma^{\rho}_{\sigma\mu}} + \Gamma^{\rho}_{\beta\mu} \Gamma^{\beta}_{\sigma\nu} - \Gamma^{\rho}_{\beta\nu} \Gamma^{\beta}_{\sigma\mu};
substitute(R4, Gamma)
evaluate(R4, g, rhsonly=True);
\end{python}
\begin{adjustwidth}{1em}{0cm}${}R^{\rho}\,_{\sigma \mu \nu} = \partial_{\mu}{\Gamma^{\rho}\,_{\sigma \nu}}-\partial_{\nu}{\Gamma^{\rho}\,_{\sigma \mu}}+\Gamma^{\rho}\,_{\beta \mu} \Gamma^{\beta}\,_{\sigma \nu}-\Gamma^{\rho}\,_{\beta \nu} \Gamma^{\beta}\,_{\sigma \mu}$\end{adjustwidth}
\begin{adjustwidth}{1em}{0cm}${}R^{\rho}\,_{\sigma \mu \nu} = \square{}_{\sigma}{}_{\nu}{}^{\rho}{}_{\mu}\brwrap{\{}{\begin{aligned}\square{}_{\varphi}{}_{\varphi}{}^{\theta}{}_{\theta}= & {\brwrap{(}{\sin{\theta}}{)}}^{2}\\[-.5ex]
\square{}_{\theta}{}_{\varphi}{}^{\varphi}{}_{\theta}= & -1\\[-.5ex]
\square{}_{\varphi}{}_{\theta}{}^{\theta}{}_{\varphi}= & -{\brwrap{(}{\sin{\theta}}{)}}^{2}\\[-.5ex]
\square{}_{\theta}{}_{\theta}{}^{\varphi}{}_{\varphi}= & 1\\[-.5ex]
\end{aligned}}{.}$
\end{adjustwidth}

The Ricci tensor $R_{\alpha\beta}$ is computed from the Riemann tensor:
\begin{python}
R2:= R_{\sigma\nu} = R^{\rho}_{\sigma\rho\nu};
substitute(R2, R4)
evaluate(R2, g, rhsonly=True);
\end{python}
\begin{adjustwidth}{1em}{0cm}${}R_{\sigma \nu} = R^{\rho}\,_{\sigma \rho \nu}$\end{adjustwidth}
\begin{adjustwidth}{1em}{0cm}${}R_{\sigma \nu} = \square{}_{\sigma}{}_{\nu}\brwrap{\{}{\begin{aligned}\square{}_{\varphi}{}_{\varphi}= & {\brwrap{(}{\sin{\theta}}{)}}^{2}\\[-.5ex]
\square{}_{\theta}{}_{\theta}= & 1\\[-.5ex]
\end{aligned}}{.}
$\end{adjustwidth}

Finally, we compute the scalar curvature $R$:
\begin{python}
R:= R = R_{\sigma\nu} g^{\sigma\nu};
substitute(R, R2)
evaluate(R, g, rhsonly=True);
\end{python}
\begin{adjustwidth}{1em}{0cm}${}R = R_{\sigma \nu} g^{\sigma \nu}$\end{adjustwidth}
\begin{adjustwidth}{1em}{0cm}${}R = 2{r}^{-2}$\end{adjustwidth}

Thus, from the user's perspective, component
computations in Cadabra~2.x are represented by the
additional property \verb|Coordinate| and several functions with the main one being \verb|evaluate()| (it is this
function that computes the components).

At the system level, component computations are
implemented in Cadabra~2.x through interaction with a
scalar computer algebra system, namely, with SymPy.

  \section{Use of Graphics in Cadabra~2.x}
\label{sec:plot}

The need for graphics in tensor-oriented computer
algebra systems is questionable. In our opinion, it is
simply not needed. However, in this case, graphics
capabilities is nothing more than an additional (side)
effect of implementing Cadabra 2.x in the framework
of the Python ecosystem. That is why graph plotting in
Cadabra 2.x is the same as in Python.

First, we need to choose a plotting library. The following example
uses the popular Matplotlib library~\cite{tosi:2009:matplotlib,vaingast:2009:matplotlib,muller:2016:python-ml}.
For numerical computations, the NumPy~\cite{idris:numpy_cookbook,oliphant:guide_numpy}
library is used.

Then, we import the modules for
\verb|matplotlib| and \verb|numpy|:
\begin{python}
import matplotlib.pyplot as plt
import numpy as np
\end{python}

Let us construct a vector field; i.e., to each point in
a space (in our case, on a plane), we assign a vector
that originates from this point. Suppose that the vector
$f$ has the form
\begin{equation}
  \label{eq:vec_field}
  f^{i} (x,y) = \mqty(\sin x \cos x\\ \cos y).
\end{equation}

Let us define a square grid on which the vector field
is evaluated. Since grid pitch along the axes is the
same, we use only the values for $x$:
\begin{python}
x = np.arange(-2*np.pi, 2*np.pi, 0.1)
u = np.sin(x)*np.cos(x)
v = np.cos(x)
uu, vv = np.meshgrid(u,v)
\end{python}

The function \verb|meshgrid| from the \verb|numpy| package
generates a rectangular grid based on two arrays (in
our case, $u$ and $v$).

Now, we construct the vector field by using the
function \verb|streamplot| from the \verb|matplotlib|
package:
\begin{python}
fig = plt.figure()
plt.streamplot(x, x, uu, vv, color='black') 
plt.title('Vector field')
plt.xlabel('x')
plt.ylabel('y')
\end{python}

The resulting plot can be both saved and displayed:
\begin{python}
fig.savefig("plot.pdf")
\end{python}
\begin{python}
display(fig)
\end{python}

\begin{figure}
  \includegraphics[width=\linewidth]{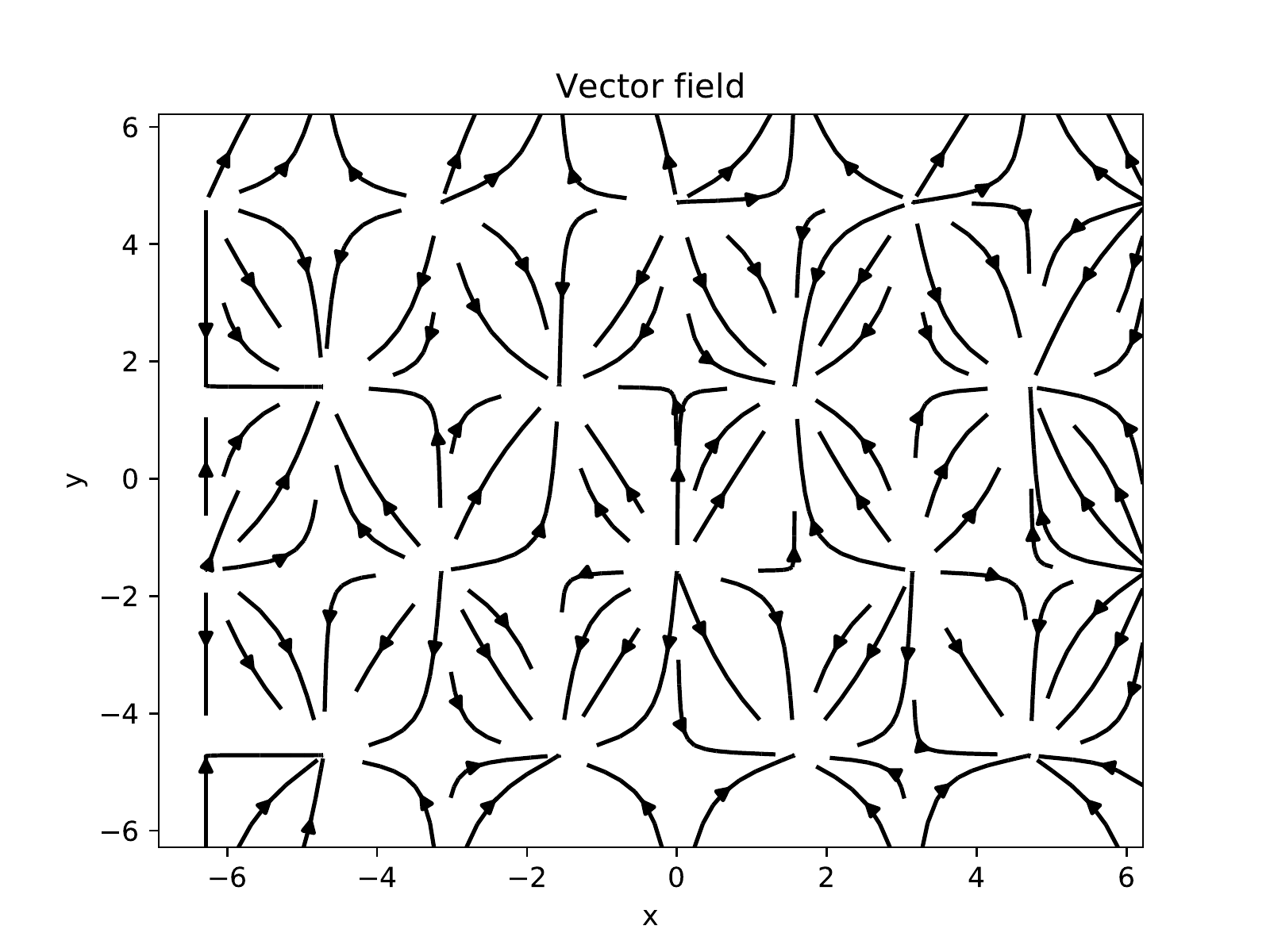}
  \caption{Image of the vector field~\eqref{eq:vec_field} constructed using matplotlib}
  \label{fig:plot}
\end{figure}

The image of the vector field is shown in Fig.~\ref{fig:plot}.

We hope that, in the future, Cadabra will include
more useful applications than the one illustrated by
this simple example.  

\section{Conclusion}
\label{sec:conclusion}

We can make the following conclusions. From the
user's perspective, the main breakthrough of Cadabra~2.x is the implementation of component computations, which allows the system to cover the whole
range of necessary tensor operations. From the developer's perspective, the main innovation is rewriting
the system by using the Python language and its entire
ecosystem. We hope that this will increase interest in
Cadabra~2.x when solving problems that involve tensor
operations.

\begin{acknowledgments}

The publication has been prepared with the support of the ``RUDN University Program 5-100'' 
and funded by Russian Foundation for Basic Research (RFBR) according to the research project 
No~16-07-00556, 18-07-00567, 18-51-18005.

\end{acknowledgments}

 \bibliographystyle{elsarticle-num}

\bibliography{bib/cadabra2/cite}

\begin{thebibliography}{10}
\def\selectlanguageifdefined#1{
\expandafter\ifx\csname date#1\endcsname\relax
\else\selectlanguage{#1}\fi}
\providecommand*{\href}[2]{{\small #2}}
\providecommand*{\url}[1]{{\small #1}}
\providecommand*{\BibUrl}[1]{\url{#1}}
\providecommand{\BibAnnote}[1]{}
\providecommand*{\BibEmph}[1]{#1}
\ProvideTextCommandDefault{\cyrdash}{\iflanguage{russian}{\hbox
  to.8em{--\hss--}}{\textemdash}}
\providecommand*{\BibDash}{\ifdim\lastskip>0pt\unskip\nobreak\hskip.2em plus
  0.1em\fi
\cyrdash\hskip.2em plus 0.1em\ignorespaces}
\renewcommand{\newblock}{\ignorespaces}

\bibitem{maccallum:2018:computer-algebra-gravity}
\selectlanguageifdefined{english}
\BibEmph{MacCallum~M. A.~H.} {Computer algebra in gravity research}~//
  \href{http://dx.doi.org/10.1007/s41114-018-0015-6}{\BibEmph{Living Reviews in
  Relativity}}. \BibDash
\newblock 2018. \BibDash dec. \BibDash
\newblock Vol.~21, no.~1. \BibDash
\newblock P.~1--93.

\bibitem{ilyin:1996:atensor}
\selectlanguageifdefined{english}
\BibEmph{Ilyin~V., Kryukov~A.} {ATENSOR — REDUCE program for tensor
  simplification}~//
  \href{http://dx.doi.org/10.1016/0010-4655(96)00060-4}{\BibEmph{Computer
  Physics Communications}}. \BibDash
\newblock 1996. \BibDash jul. \BibDash
\newblock Vol.~96, no.~1. \BibDash
\newblock P.~36--52.

\bibitem{gomez-lobo:2012:spinors-xact}
\selectlanguageifdefined{english}
\BibEmph{G{\'{o}}mez-Lobo~A. G.~P., Mart{\'{i}}n-Garc{\'{i}}a~J.~M.} {Spinors:
  A Mathematica package for doing spinor calculus in General Relativity}~//
  \href{http://dx.doi.org/10.1016/j.cpc.2012.04.024}{\BibEmph{Computer Physics
  Communications}}. \BibDash
\newblock 2012. \BibDash
\newblock Vol. 183, no.~10. \BibDash
\newblock P.~2214--2225. \BibDash
\newblock arXiv~: 1110.2662.

\bibitem{maccallum:2002:ca-gr}
\selectlanguageifdefined{english}
\BibEmph{MacCallum~M.} {Computer Algebra in General Relativity}~//
  \href{http://dx.doi.org/10.1142/S0217751X02011643}{\BibEmph{International
  Journal of Modern Physics A}}. \BibDash
\newblock 2002. \BibDash
\newblock Vol.~17, no.~20. \BibDash
\newblock P.~2707--2710.

\bibitem{bolotin:2013:redberry}
\selectlanguageifdefined{english}
\BibEmph{Bolotin~D.~A., Poslavsky~S.~V.} {Introduction to Redberry: the
  Computer Algebra System Designed for Tensor Manipulation}. \BibDash
\newblock 2015. \BibDash
\newblock P.~1--27. \BibDash
\newblock arXiv~: 1302.1219.

\bibitem{poslavsky:2015:redberry}
\selectlanguageifdefined{english}
\BibEmph{Poslavsky~S., Bolotin~D.} {Redberry: a computer algebra system
  designed for tensor manipulation}~//
  \href{http://dx.doi.org/10.1088/1742-6596/608/1/012060}{\BibEmph{Journal of
  Physics: Conference Series}}. \BibDash
\newblock 2015. \BibDash may. \BibDash
\newblock Vol. 608, no.~1. \BibDash
\newblock P.~012060. \BibDash
\newblock arXiv~: 1302.1219.

\bibitem{fliegner:1999:form}
\selectlanguageifdefined{english}
\BibEmph{Fliegner~D., Retey~A., Vermaseren~J. a.~M.} {Parallelizing the
  Symbolic Manipulation Program FORM Part I: Workstation Clusters and Message
  Passing}. \BibDash
\newblock 2000. \BibDash
\newblock arXiv~: hep-ph/0007221.

\bibitem{heck:2000:formbook}
\selectlanguageifdefined{english}
\BibEmph{Heck~A.} {FORM for Pedestrians}. \BibDash
\newblock 2000.

\bibitem{tung:2005:form}
\selectlanguageifdefined{english}
\BibEmph{Tung~M.~M.} {FORM Matters: Fast Symbolic Computation under UNIX}~//
  \href{http://dx.doi.org/10.1016/j.camwa.2004.07.023}{\BibEmph{Computers and
  Mathematics with Applications}}. \BibDash
\newblock 2005. \BibDash
\newblock Vol.~49, no. 7-8. \BibDash
\newblock P.~1127--1137. \BibDash
\newblock arXiv~: cs/0409048.

\bibitem{peeters:2007:cadabra}
\selectlanguageifdefined{english}
\BibEmph{Peeters~K.} {Introducing Cadabra: a symbolic computer algebra system
  for field theory problems}. \BibDash
\newblock 2007. \BibDash
\newblock arXiv~: hep-th/0701238.

\bibitem{peeters:2007:cpc:cadabra}
\selectlanguageifdefined{english}
\BibEmph{Peeters~K.} {Cadabra: a field-theory motivated symbolic computer
  algebra system}~//
  \href{http://dx.doi.org/10.1016/j.cpc.2007.01.003}{\BibEmph{Computer Physics
  Communications}}. \BibDash
\newblock 2007. \BibDash
\newblock Vol. 176, no.~8. \BibDash
\newblock P.~550--558. \BibDash
\newblock arXiv~: cs/0608005.

\bibitem{brewin:2010:cpc:tensor-cadabra}
\selectlanguageifdefined{english}
\BibEmph{Brewin~L.} {A Brief Introduction to Cadabra: A Tool for Tensor
  Computations in General Relativity}~//
  \href{http://dx.doi.org/10.1016/j.cpc.2009.10.020}{\BibEmph{Computer Physics
  Communications}}. \BibDash
\newblock 2010. \BibDash mar. \BibDash
\newblock Vol. 181, no.~3. \BibDash
\newblock P.~489--498. \BibDash
\newblock arXiv~: 0903.2085.

\bibitem{kulyabov:2009:nucleilett:cadabra}
\selectlanguageifdefined{english}
\BibEmph{Sevastianov~L.~A., Kulyabov~D.~S., Kokotchikova~M.~G.} {An Application
  of Computer Algebra System Cadabra to Scientific Problems of Physics}~//
  \href{http://dx.doi.org/10.1134/S1547477109070073}{\BibEmph{Physics of
  Particles and Nuclei Letters}}. \BibDash
\newblock 2009. \BibDash
\newblock Vol.~6, no.~7. \BibDash
\newblock P.~530--534.

\bibitem{kulyabov:2013:springer:cadabra}
\selectlanguageifdefined{english}
\BibEmph{Korol'kova~A.~V., Kulyabov~D.~S., Sevast'yanov~L.~A.} {Tensor
  Computations in Computer Algebra Systems}~//
  \href{http://dx.doi.org/10.1134/S0361768813030031}{\BibEmph{Programming and
  Computer Software}}. \BibDash
\newblock 2013. \BibDash
\newblock Vol.~39, no.~3. \BibDash
\newblock P.~135--142. \BibDash
\newblock arXiv~: 1402.6635.

\bibitem{kulyabov:2016:pcs}
\selectlanguageifdefined{english}
\BibEmph{Kulyabov~D.~S.} {Using two Types of Computer Algebra Systems to Solve
  Maxwell Optics Problems}~//
  \href{http://dx.doi.org/10.1134/S0361768816020043}{\BibEmph{Programming and
  Computer Software}}. \BibDash
\newblock 2016. \BibDash
\newblock Vol.~42, no.~2. \BibDash
\newblock P.~77--83. \BibDash
\newblock arXiv~: 1605.00832.

\bibitem{leeuwen:1992:lie}
\selectlanguageifdefined{english}
\BibEmph{Leeuwen~M. A. A.~v., Cohen~A.~M., Lisser~B.} {LiE: A package for Lie
  group computations}. \BibDash
\newblock Amsterdam~: Computer Algebra Nederland, 1992.

\bibitem{oliphant:2007:scipy}
\selectlanguageifdefined{english}
\BibEmph{Oliphant~T.~E.} {Python for Scientific Computing}~//
  \href{http://dx.doi.org/10.1109/MCSE.2007.58}{\BibEmph{Computing in Science
  and Engineering}}. \BibDash
\newblock 2007. \BibDash
\newblock Vol.~9, no.~3. \BibDash
\newblock P.~10--20.

\bibitem{ll:2::ru}
\selectlanguageifdefined{russian}
\BibEmph{Ландау~Л.~Д., Лифшиц~Е.~М.} {Теория поля}.
  Теоретическая физика. Т. II. \BibDash
\newblock 8-е {изд.} \BibDash
\newblock М.~: Физматлит, 2012. \BibDash
\newblock 536~{с.}

\bibitem{perez:2007:ipython}
\selectlanguageifdefined{english}
\BibEmph{Perez~F., Granger~B.~E.} {IPython: A System for Interactive Scientific
  Computing}~//
  \href{http://dx.doi.org/10.1109/MCSE.2007.53}{\BibEmph{Computing in Science
  and Engineering}}. \BibDash
\newblock 2007. \BibDash
\newblock Vol.~9, no.~3. \BibDash
\newblock P.~21--29.

\bibitem{rfc:7763}
\selectlanguageifdefined{english}
\href{http://dx.doi.org/10.17487/RFC7763}{{The text/markdown Media Type}}~:
  RFC~/ RFC Editor~; Executor: S.~Leonard~: 2016. \BibDash mar.

\bibitem{lamy:sympy_starter}
\selectlanguageifdefined{english}
\BibEmph{Lamy~R.} {Instant SymPy Starter}. \BibDash
\newblock Packt Publishing, 2013. \BibDash
\newblock 52~p.

\bibitem{mtw:1::ru}
\selectlanguageifdefined{russian}
\BibEmph{Мизнер~Ч., Торн~К., Уилер~Д.~А.}
  {Гравитация}. \BibDash
\newblock Мир {изд.} \BibDash
\newblock М., 1977. \BibDash
\newblock Т.~1. \BibDash
\newblock 474~{с.} \BibAnnote{Автор: Чарльз Минзер, Кип
  Торн, Джон Уилер Жанр: Учебное пособие
  Язык оригинала: Английский
  Оригинал издан: 1977 Переводчик: Перевод с
  английского М. М. Баско (Т. 1), A. А.
  Рузмайкина (Т. 2) и A. Г. Полнарева (Т. 3) под
  редакцией В. Б. Брагинского и И. Д. Новикова
  Издатель: М. — Издательство «Мир».
  Редакция литературы по физике Выпуск: 1973
  Страниц: 474 (том 1), 525 (том 2), 510 (том 3).}

\bibitem{tosi:2009:matplotlib}
\selectlanguageifdefined{english}
\BibEmph{Tosi~S.} {Matplotlib for Python Developers}. \BibDash
\newblock Packt Publishing, 2009. \BibDash
\newblock 308~p.

\bibitem{vaingast:2009:matplotlib}
\selectlanguageifdefined{english}
\BibEmph{Vaingast~S.} {Beginning Python visualization: crafting visual
  transformation scripts}. \BibDash
\newblock Springer, 2009. \BibDash
\newblock 384~p.

\bibitem{muller:2016:python-ml}
\selectlanguageifdefined{english}
\BibEmph{M{\"{u}}ller~A.~C., Guido~S.} {Introduction to Machine Learning with
  Python: A Guide for Data Scientists}. \BibDash
\newblock O'Reilly Media, 2016. \BibDash
\newblock 285~p.

\bibitem{idris:numpy_cookbook}
\selectlanguageifdefined{english}
\BibEmph{Idris~I.} {NumPy Cookbook}. \BibDash
\newblock Packt Publishing, 2012. \BibDash
\newblock 226~p.

\bibitem{oliphant:guide_numpy}
\selectlanguageifdefined{english}
\BibEmph{Oliphant~T.~E.} {Guide to NumPy}. \BibDash
\newblock 2 edition. \BibDash
\newblock CreateSpace Independent Publishing Platform, 2015. \BibDash
\newblock 364~p.

\end{thebibliography}

\end{document}


\graphicspath{{image/cadabra2/ru/}{image/cadabra2/}}

\title{Новые возможности второй версии пакета компьютерной алгебры Cadabra}

\author{Д. С. Кулябов}
\email{kulyabov-ds@rudn.ru}
\affiliation{Кафедра прикладной информатики и теории вероятностей,\\
  Российский университет дружбы народов,\\
  117198, Москва, ул. Миклухо-Маклая, д. 6}
\affiliation{Лаборатория информационных технологий,\\
  Объединённый институт ядерных исследований,\\
  ул. Жолио-Кюри 6, Дубна, Московская область, Россия, 141980}

\author{А. В. Королькова}
\email{korolkova-av@rudn.ru}
\affiliation{Кафедра прикладной информатики и теории вероятностей,\\
  Российский университет дружбы народов,\\
  117198, Москва, ул. Миклухо-Маклая, д. 6}

\author{Л. А. Севастьянов}
\email{sevastianov-la@rudn.ru}
\affiliation{Кафедра прикладной информатики и теории вероятностей,\\
  Российский университет дружбы народов,\\
  117198, Москва, ул. Миклухо-Маклая, д. 6}
\affiliation{Лаборатория теоретической физики,\\
  Объединённый институт ядерных исследований,\\
  ул. Жолио-Кюри 6, Дубна, Московская область, Россия, 141980}

\begin{abstract}

  В определённых научных областях есть потребность использования
  операций над тензорами. Для облегчения кропотливой работы над
  тензорными расчётами можно использовать системы компьютерной
  алгебры.  В качестве основной системы тензорной компьютерной алгебры
  авторами данной работы в своих научных исследованиях уже несколько
  лет используется система Cadabra.
  Недавно вышла работоспособная вторая версия этой программы. В ней
  сделан ряд улучшений, которые можно позиционировать 
  как революционные. Наиболее яркие улучшения касаются реализации
  компонентных тензорных операций и смены идеологии программной
  реализации системы по сравнению с первой версией Cadabra. В данной
  статье даётся краткий обзор ключевых улучшений в системе Cadabra.

\end{abstract}

  \keywords{системы компьютерной алгебры, тензорные вычисления,
    компонентные вычисления, python, sympy}

\maketitle

\section{Введение}
\label{sec:intro}

  Системы компьютерной алгебры с поддержкой операций над тензорами
  можно условно разделить на три
  группы~\cite{maccallum:2018:computer-algebra-gravity}. К первой
  группе относят универсальные системы тензорных вычислений, которые
  создавались в основном для применения в задачах общей теории
  относительности. В таких системах упор делается на вычисление
  характерных величин римановой геометрии (символы Кристоффеля, тензор
  Римана)~\cite{ilyin:1996:atensor,gomez-lobo:2012:spinors-xact,maccallum:2002:ca-gr}. Ко
  второй группе относят системы тензорной компьютерной алгебры для
  расчётов в квантовой теории
  поля~\cite{bolotin:2013:redberry,poslavsky:2015:redberry,fliegner:1999:form,heck:2000:formbook,tung:2005:form}. В
  них упор делается на вычисления с использованием дираковских
  спиноров, а также используются простые симметрии.  Третья группа
  систем тензорной компьютерной алгебры создавалась скорее как
  фреймворк для произвольных тензорных расчётов. В них меньше готовых
  шаблонов, но есть более выразительные средства для конструирования
  новых объектов. Именно к этой группе и относится система Cadabra,
  особенностям реализации которой посвящена данная работа.
  

  Система Cadabra на данный момент имеет две версии.
  В первой версии Cadabra 
  (далее Cadabra1) есть существенные отличия  
  от других систем тензорной компьютерной алгебры, в частности:
\begin{itemize}
\item использование \TeX{}-нотации при записи выражений;
\item удобство определения произвольного типа тензоров;
\item использование диаграмм Юнга для описания симметрийных свойств тензоров.
\end{itemize}

Тем не менее 
Cadabra1 имеет и существенные недостатки. В частности, в этой версии
отсутствует возможность компонентных тензорных вычислений. Кроме того,
монолитная и косная программная структура самой системы не даёт
возможности надеяться на реализацию компонентных вычислений в
ближайшем будущем. Однако, революционный шаг, совершённый при
реализации второй версии системы (далее Cadabra2), а именно
использование экосистемы Python, позволило решить указанные
проблемы. На наш взгляд именно комбинация <<компонентные вычисления +
экосистема Python>>, реализованная во второй версии, является
революционным улучшением системы Cadabra.

К сожалению, документация по системе Cadabra оставляет желать
лучшего. На сайте самой системы (\url{https://cadabra.science/})
представлен достаточно небольшой набор примеров. Более существенную
информацию можно получить из специализированных исследовательских
статей, в которых можно встретить возможное применение системы
Cadabra~\cite{peeters:2007:cadabra,peeters:2007:cpc:cadabra,brewin:2010:cpc:tensor-cadabra,kulyabov:2009:nucleilett:cadabra,kulyabov:2013:springer:cadabra,kulyabov:2016:pcs}.

  В данной работе предлагается небольшой обзор новых возможностей
  второй версии пакета компьютерной алгебры Cadabra. Структура статьи
  следующая.  В~разделе~\ref{sec:realisation} рассмотрены особенности
  реализации первой и второй версий Cadabra.  В
  разделе~\ref{sec:syntax} приведено описание процесса варьирования
  действия для электромагнитного поля без источников с целью сделать
  краткий обзор синтаксиса Cadabra2. Если читатель хочет сравнить
  синтаксис Cadabra2 с синтаксисом Cadabra1, то мы рекомендуем
  ознакомится с более ранними
  работами~\cite{kulyabov:2009:nucleilett:cadabra,kulyabov:2013:springer:cadabra,kulyabov:2016:pcs}.
  В разделе~\ref{sec:sympy} рассмотрена одна из важнейших особенностей
  Cadabra2~--- прозрачное взаимодействие с универсальной скалярной
  системой компьютерной алгебры SymPy. В разделе~\ref{sec:coord}
  описывается основное, на наш взгляд, нововведение в Cadabra2~---
  компонентные вычисления, рассмотренные на примере нахождения
  основных величин для общей теории относительности, а именно
  связности Кристоффеля и разных кривизн.  В завершении статьи, в
  разделе~\ref{sec:plot} приводится пример, демонстрирующий работу с
  графикой в 
  Cadabra2. 

  Заметим, что при реализации примеров были использованы элементы кода
  из документации по Cadabra2
  (\url{https://cadabra.science/tutorials.html}).

\section{Особенности реализации Cadabra1 и Cadabra2}

\label{sec:realisation}

Каждая система компьютерной алгебры имеет свою специфику в реализации.
Можно выделить несколько уровней при реализации:
\begin{itemize}
\item нотация;
\item язык манипуляции;
\item язык реализации;
\item язык расширений.
\end{itemize}

Не у каждой системы компьютерной алгебры можно выделить все эти
уровни. Так, у большинства систем нотация строится на основе языка
манипуляции. 

У пакета Cadabra2, также как и у пакета Cadabra1 нотация построена на
основе \TeX{}-нотации (правильнее сказать, \TeX{}-подобной нотации).
Для этого выделяется некоторое подмножество символов \TeX{}, как то
буквы разных алфавитов, символы производных, интеграла и так
далее. Главное, что индексы обозначаются с помощью символов \verb|_| и
\verb|^|, как привычно пользователям системы \TeX{}.

Язык манипуляций служит собственно для работы в системе.  Синтаксис
Cadabra1 достаточно простой и скорее приспособлен для облегчения
лексического и синтаксического разбора текста программы, чем для
удобства пользователя.
Такой подход к конструированию языков использовали в ранние годы
развития вычислительной техники (например, в языках Shell, Perl). В
Cadabra2 произошёл качественный скачок --- система перешла на язык
Python. Соответственно и все операции в Cadabra2 выполняются в
Python-подобном синтаксисе.

Реализация системы Cadabra1 выполнена на языке C++ с использованием
системы компьютерной алгебры LiE~\cite{leeuwen:1992:lie} (на данный
момент компиляция системы LiE вызывает некоторые трудности). Основное
предназначение --- работа с группами Ли, на которых основаны операции
с симметриями тензоров.  Следует
заметить, что  фактически вся система была реализована одним
человеком. Это большой труд. Но наличие только одного автора и
монолитная программная структура системы  вызвала
тревогу за дальнейшую судьбу всей системы.

В версии Cadabra2 автор принципиально изменил свой подход к структуре
системы. Он взял направление на интеграцию её с экосистемой
Python. Система всё также написана на языке C++, но в качестве
языка-клея применяется Python, который также используется в качестве
языка манипуляций.  При этом Python же может быть использован как язык
написания расширений.

Инфраструктура Python открывает доступ к огромному количеству научных
библиотек, в том числе и к проекту
SciPy~\cite{oliphant:2007:scipy}. Это позволяет использовать
библиотеки SciPy бесшовно, прозрачно для пользователя. Именно это и
позволило на наш взгляд сделать системе Cadabra2 революционный шаг.

\section{Элементы синтаксиса Cadabra2}
\label{sec:syntax}

Напомним некоторые элементы синтаксиса Cadabra1 и Cadabra2. В
качестве примера рассмотрим 
получение уравнения Максвелла без источников~\cite{ll:2::ru}:
\begin{equation}
  \label{eq:F_mn}
  \partial_{\mu} F^{\mu \nu} = 0,
\end{equation}
варьируя действие
\begin{equation}
  \label{eq:S-maxwell}
  S = - \frac{1}{4} \int F_{\mu\nu} F^{\mu\nu} \dd{x}.
\end{equation}

При этом тензор Максвелла $F_{\mu \nu}$ представим через векторный потенциал:
\begin{equation}
  \label{eq:F=A-A}
  F_{\mu\nu} = \partial_{\mu}{A_{\nu}} - \partial_{\nu}{A_{\mu}}.
\end{equation}

Напомним, что обе версии Cadabra используют нотацию \TeX{}. Во второй
версии появилась возможность использовать функционал языка Python, в
частности, можно задавать функции прямо в тексте программы, написанной
на Cadabra2. 
Заметим, что также как и Cadabra1, Cadabra2 по
умолчанию практически не обрабатывает получающиеся выражения (разве
что приводит подобные), эта обработка отдана на откуп самому
пользователю. 
Если требуется применить какой-то набор правил ко всем используемым
выражениям, то можно задать функцию \verb|post_process|, выполняемую
после каждой операции (фактически, \verb|post_process| является просто
функцией Python):
\begin{python}
def post_process(ex):
    sort_product(ex)
    canonicalise(ex)
    collect_terms(ex)
\end{python}

В экстремальном случае эту функцию можно сделать пустой, убрав таким
образом любую обработку выражений.

Интерфейс Cadabra2 выполнен по идеологии блокнота, то есть наряду с
программным кодом можно указывать и пояснения. Правда, в отличие от
iPython~\cite{perez:2007:ipython} текстовая часть пишется не в
синтаксисе Markdown~\cite{rfc:7763}, а в синтаксисе \LaTeX{}.

Объекту Cadabra2 можно приписать свойство, имеющее 
в свою очередь
некоторый набор настроек. Естественно, раз речь идёт о
тензорах, наиболее полезное свойство~--- \verb|Indices|. Настройка
\verb|position=free| даёт возможность системе самой поднимать и
опускать индексы:
\begin{python}
{\mu,\nu,\rho}::Indices(position=free).
x::Coordinate.
\partial{#}::Derivative.
\end{python}

Здесь знак \verb|#| является шаблоном подстановки. Знак точки в конце
выражения подавляет вывод (как часто принято в системах компьютерной
алгебры).

Для работы с абстрактными индексами необходимо учитывать симметрийные
свойства тензоров. Кроме того, при дифференцировании и интегрировании
необходимо учитывать зависимость объектов от координат:
\begin{python}
F_{\mu\nu}::AntiSymmetric;
F_{\mu\nu}::Depends(x).
A_{\mu}::Depends(x,\partial{#}).
\delta{#}::Accent;
\end{python}
\begin{adjustwidth}{1em}{0cm}${}\text{Attached property AntiSymmetric to~}F_{\mu \nu}.$\end{adjustwidth}
\begin{adjustwidth}{1em}{0cm}${}\text{Attached property Accent to~}\delta{\#}.$\end{adjustwidth}

В данном случае знак вариации $\delta$ 
рассматривается как модификатор, а не как объект со свойством
\verb|Derivative|, и не вносит никакой дополнительной вычислительной
семантики.

В рассматриваемом нами примере 
$F_{\mu\nu}$ является тензором Максвелла. Распишем его
через векторный потенциал $A_{\mu}$~\eqref{eq:F=A-A}:
\begin{python}
F:= F_{\mu\nu} = \partial_{\mu}{A_{\nu}} - \partial_{\nu}{A_{\mu}};
\end{python}
\begin{adjustwidth}{1em}{0cm}${}F_{\mu \nu} = \partial_{\mu}{A_{\nu}}-\partial_{\nu}{A_{\mu}}$\end{adjustwidth}

Здесь знаком $:=$ задаётся метка строки (в нашем случае это $F$).
Метка именует выражение для удобства ссылки на него в последствии.

Далее зададим действие для электромагнитного
поля~\eqref{eq:S-maxwell}:
\begin{python}
S:= -1/4 \int{ F_{\mu\nu} F^{\mu\nu} }{x};
\end{python}
\begin{adjustwidth}{1em}{0cm}${} - \frac{1}{4}\int F^{\mu \nu} F_{\mu \nu}\,\,{\rm d}x$\end{adjustwidth}

Путём подстановки перепишем действие через векторный потенциал $A_{\mu}$:
\begin{python}
substitute(S, F);
\end{python}
\begin{adjustwidth}{1em}{0cm}${} - \frac{1}{4}\int \brwrap{(}{\partial_{\mu}{A_{\nu}}-\partial_{\nu}{A_{\mu}}}{)} \brwrap{(}{\partial^{\mu}{A^{\nu}}-\partial^{\nu}{A^{\mu}}}{)}\,\,{\rm d}x$\end{adjustwidth}
После этого посмотрим вид действия:
\begin{python}
S;
\end{python}
\begin{adjustwidth}{1em}{0cm}${} - \frac{1}{4}\int \brwrap{(}{\partial_{\mu}{A_{\nu}}-\partial_{\nu}{A_{\mu}}}{)} \brwrap{(}{\partial^{\mu}{A^{\nu}}-\partial^{\nu}{A^{\mu}}}{)}\,\,{\rm d}x$\end{adjustwidth}

Значение действия в этом выражении не такое, какое мы задали
изначально. Оно стало таким, каким мы его получили после последних
вычислений.  И это немного необычно. Дело в том, что большинство
систем компьютерной алгебры реализованы на функциональных языках
программирования или придерживаются функциональной парадигмы, в рамках
которой 
переменные обладают свойством иммутабельности. В данном же случае
метка ведёт себя как переменная в императивных языках программирования
(собственно, Python и является императивным языком). Это делает работу
в Cadabra2 необходимо линейной: нельзя произвольно передвигаться по
блокноту и производить вычисления в произвольном месте.

Проварьируем действие: 
\begin{python}
vary(S, $A_{\mu} -> \delta{A_{\mu}}$);
\end{python}
\begin{adjustwidth}{1em}{0cm}${} - \frac{1}{4}\int \brwrap{(}{\brwrap{(}{\partial^{\mu}{A^{\nu}}-\partial^{\nu}{A^{\mu}}}{)} \brwrap{(}{\partial_{\mu}{\delta{A_{\nu}}}-\partial_{\nu}{\delta{A_{\mu}}}}{)}+\brwrap{(}{\partial_{\mu}{A_{\nu}}-\partial_{\nu}{A_{\mu}}}{)} \brwrap{(}{\partial^{\mu}{\delta{A^{\nu}}}-\partial^{\nu}{\delta{A^{\mu}}}}{)}}{)}\,\,{\rm d}x$\end{adjustwidth}

Здесь мы видим пример использования не метки, а непосредственно
выражения. Для этого мы окружаем выражение символами доллара (\$), как
делается в обычном \TeX{}'е.

Далее раскроем произведения и приведём подобные:
\begin{python}
distribute(S);
\end{python}
\begin{adjustwidth}{1em}{0cm}${} - \frac{1}{4}\int \brwrap{(}{4\partial^{\mu}{A^{\nu}} \partial_{\mu}{\delta{A_{\nu}}}-4\partial^{\mu}{A^{\nu}} \partial_{\nu}{\delta{A_{\mu}}}}{)}\,\,{\rm d}x$\end{adjustwidth}

Затем проинтегрируем по частям: 
\begin{python}
integrate_by_parts(S, $\delta{A_{\mu}}$);
\end{python}
\begin{adjustwidth}{1em}{0cm}${} - \frac{1}{4}\int \brwrap{(}{-4\delta{A^{\mu}} \partial^{\nu}\brwrap{(}{\partial_{\nu}{A_{\mu}}}{)}+4\delta{A^{\mu}} \partial^{\nu}\brwrap{(}{\partial_{\mu}{A_{\nu}}}{)}}{)}\,\,{\rm d}x$\end{adjustwidth}

Интегрирование по частям в данном случае является достаточно
формальным действием, использующим только свойство \verb|Derivative|
объекта.

Ещё раз выполним подстановку, а затем раскроем произведения и приведём
подобные:
\begin{python}
substitute(_, $\partial_{\mu}{A_{\nu}} -> 1/2 \partial_{\mu}{A_{\nu}} + 1/2 F_{\mu\nu} + 1/2 \partial_{\nu}{ A_{\mu}}$);
\end{python}
\begin{adjustwidth}{1em}{0cm}${} - \frac{1}{4}\int \brwrap{(}{-4\delta{A^{\mu}} \partial^{\nu}\brwrap{(}{\frac{1}{2}\partial_{\nu}{A_{\mu}} - \frac{1}{2}F_{\mu \nu}+\frac{1}{2}\partial_{\mu}{A_{\nu}}}{)}+4\delta{A^{\mu}} \partial^{\nu}\brwrap{(}{\frac{1}{2}\partial_{\mu}{A_{\nu}}+\frac{1}{2}F_{\mu \nu}+\frac{1}{2}\partial_{\nu}{A_{\mu}}}{)}}{)}\,\,{\rm d}x$\end{adjustwidth}
\begin{python}
distribute(_);
\end{python}
\begin{adjustwidth}{1em}{0cm}${}-\int \delta{A^{\mu}}
  \partial^{\nu}{F_{\mu \nu}}\,\,{\rm d}x$\end{adjustwidth}

Здесь специальная метка \verb|_| указывает на предыдущее выражение.

В результате получим искомое уравнение Максвелла~\eqref{eq:F_mn}:
\begin{equation}
  \partial^\nu F_{\mu\nu}=0
\end{equation}

Таким образом показано, что синтаксис языка манипуляции выражениями в
системе Cadabra2 базируется на синтаксисе языка Python и он более
привычен, чем синтаксис языка Cadabra1.

\section{Взаимодействие Cadabra с SymPy}
\label{sec:sympy}

Тензорные системы компьютерной алгебры имеют в своём арсенале
достаточно небольшое число поддерживаемых операций. Их достаточно для
основных манипуляций с тензорами в формализме абстрактных индексов и в
безындексном формализме. Но для многих операций (например, для
полноценной реализации компонентных вычислений) требуются возможность
использования скалярных операций. 
Если тензорная система компьютерной алгебры реализована в рамках
универсальной системы компьютерной алгебры, то проблем не
возникает. Однако Cadabra является отдельной системой. В Cadabra1
реализован, хотя и неудобный в применении, механизм для связи с
универсальной системой компьютерной алгебры Maxima
--- 
создалось впечатление, что этот механизм реализован только как
<<доказательство концепции>>.

В Cadabra2 связь с универсальной системой компьютерной алгебры
реализована через SymPy~\cite{lamy:sympy_starter}. Причём эта связь
бесшовная: для пользователя использование данного механизма проходит
незаметно. Впрочем, это один из результатов реализации системы
Cadabra2 на языке Python.

Продемонстрируем применение SymPy в Cadabra2 на примере, в котором нам
необходимо вычислить интеграл:
\begin{equation}
  \label{eq:int-sympy}
  \int \frac{1}{x} \dd{x}.
\end{equation}

Основная функция явного вызова SymPy~--- \verb|map_sympy()|. Эта
функция имеет побочное действие: она изменяет значение
аргумента. Впрочем, как мы уже обсуждали, отсутствие иммутабельности
является особенностью системы Cadabra2.  Рассмотрим простейший вызов
этой функции:
\begin{python}
ex := \int{1/x}{x};
\end{python}
\begin{adjustwidth}{1em}{0cm}${}\int {x}^{-1}\,\,{\rm d}x$\end{adjustwidth}
\begin{python}
map_sympy(_);
\end{python}
\begin{adjustwidth}{1em}{0cm}${}\log\brwrap{(}{x}{)}$\end{adjustwidth}

Для того, чтобы убедиться в наличии побочного действия, посмотрим
текущее значение выражения \verb|ex|:
\begin{python}
ex;
\end{python}
\begin{adjustwidth}{1em}{0cm}${}\log\brwrap{(}{x}{)}$\end{adjustwidth}

Мы увидим, что значение \verb|ex| изменилось.

Можно передать значение выражения не просто в систему SymPy, а вызвать
для его обработки конкретную функцию. Например, в рассматриваемом нами
случае можно вызвать функцию \verb|integrate| среды SymPy:
\begin{python}
ex := 1/x;
\end{python}
\begin{adjustwidth}{1em}{0cm}${}{x}^{-1}$\end{adjustwidth}

Вторым аргументом функции \verb|map_sympy()| идёт имя конкретной
функции SymPy:
\begin{python}
map_sympy(ex,"integrate");
\end{python}
\begin{adjustwidth}{1em}{0cm}${}\log\brwrap{(}{x}{)}$\end{adjustwidth}
\begin{python}
ex;
\end{python}
\begin{adjustwidth}{1em}{0cm}${}\log\brwrap{(}{x}{)}$\end{adjustwidth}

Здесь ещё раз убеждаемся в наличии побочного действия функции \verb|map_sympy()|.

В рамках языка Python есть несколько вариантов выполнения одного и
того же действия. Естественно, эта возможность есть и в
Cadabra2. Рассмотрим следующие варианты исключительно как
потенциальные возможности, а не как руководство к действию.

Возможно использование метода класса \verb|_sympy_()|: 
\begin{python}
ex := \int{1/x}{x};
\end{python}
\begin{adjustwidth}{1em}{0cm}${}\int {x}^{-1}\,\,{\rm d}x$\end{adjustwidth}

Воспринимая метку \verb|ex| как объект, вызовем метод \verb|_sympy_()|:
\begin{python}
ex._sympy_();
\end{python}
\begin{adjustwidth}{1em}{0cm}${}\log{\left (x \right)}$\end{adjustwidth}

Проверим состояние среды:
\begin{python}
ex;
\end{python}
\begin{adjustwidth}{1em}{0cm}${}\int {x}^{-1}\,\,{\rm d}x$\end{adjustwidth}

Видим, что состояние среды не изменилось, то есть метод класса \verb|_sympy_()|
не имеет побочного действия.

Кроме того, можно использовать функцию \verb|sympy| с методом,
соответствующим вызываемой функции среды SymPy:
\begin{python}
ex := 1/x;
\end{python}
\begin{adjustwidth}{1em}{0cm}${}{x}^{-1}$\end{adjustwidth}

Вызовем функцию \verb|sympy| с методом \verb|integrate|:
\begin{python}
sympy.integrate(ex);
\end{python}
\begin{adjustwidth}{1em}{0cm}${}\log{\left (x
    \right)}$\end{adjustwidth}

Обратим внимание, что данная функция всегда требует указания
конкретного метода, поэтому мы не смогли бы применить её в предыдущем
случае.

Опять проверим состояние среды:
\begin{python}
ex;
\end{python}
\begin{adjustwidth}{1em}{0cm}${}{x}^{-1}$\end{adjustwidth}

Убеждаемся, что функция \verb|sympy| не обладает побочным действием.

Базируясь на приведённых выше примерах, можно сделать вывод, что 
взаимодействие с SymPy в Cadabra2 реализовано достаточно элегантным
образом. Но основное достоинство этой операции заключается на наш
взгляд в её тесной интеграции с системой, например, для реализации
компонентных вычислений (см. раздел~\ref{sec:coord}).

\section{Компонентные вычисления в Cadabra2}
\label{sec:coord}

В качестве примера компонентных тензорных операций получим кривизну $R$ на
сфере $S^{2}$ радиуса $r$:
\begin{equation}
  g_{\alpha\beta} = \diag\qty(r^2,r^2 \sin^2 \vartheta).
\end{equation}

Для этого получим значения для символов Кристоффеля
$\Gamma^{\alpha}_{\mu\nu}$, тензора Римана
$R^{\alpha}\,_{\beta \mu \nu}$ и тензора Риччи $R_{\alpha\beta}$~\cite{mtw:1::ru,ll:2::ru}.

Зададим координаты и набор меток для индексов. При этом укажем,
какие значения данные метки принимают:
\begin{python}
{\theta, \varphi}::Coordinate;
{\alpha, \beta, \gamma, \delta, \rho, \sigma, \mu, \nu, \lambda}::Indices(values={\varphi, \theta}, position=fixed);
\partial{#}::PartialDerivative;
\end{python}
\begin{adjustwidth}{1em}{0cm}${}\text{Attached property Coordinate to~}\brwrap{[}{\theta,~\discretionary{}{}{} \varphi}{]}.$\end{adjustwidth}
\begin{adjustwidth}{1em}{0cm}${}\text{Attached property Indices(position=fixed) to~}\brwrap{[}{\alpha,~\discretionary{}{}{} \beta,~\discretionary{}{}{} \gamma,~\discretionary{}{}{} \delta,~\discretionary{}{}{} \rho,~\discretionary{}{}{} \sigma,~\discretionary{}{}{} \mu,~\discretionary{}{}{} \nu,~\discretionary{}{}{} \lambda}{]}.$\end{adjustwidth}
\begin{adjustwidth}{1em}{0cm}${}\text{Attached property PartialDerivative to~}\partial{\#}.$\end{adjustwidth}

Затем зададим тензор $g$ со свойством \verb|Metric| и аналогичным
образом зададим  обратную метрику:
\begin{python}
g_{\alpha\beta}::Metric.
g^{\alpha\beta}::InverseMetric.
\end{python}

В данном случае достаточно задать компоненты для самой метрики. Компоненты для
обратной метрики рассчитываются с помощью функции \verb|complete|:
\begin{python}
g:={ g_{\theta\theta} = r**2, g_{\varphi\varphi} = r**2 \sin(\theta)**2 }.
complete(g, $g^{\alpha\beta}$);
\end{python}
\begin{adjustwidth}{1em}{0cm}${}\brwrap{[}{g_{\theta \theta} = {r}^{2},~\discretionary{}{}{} g_{\varphi \varphi} = {r}^{2} {\brwrap{(}{\sin{\theta}}{)}}^{2},~\discretionary{}{}{} g^{\varphi \varphi} = {\brwrap{(}{{r}^{2} {\brwrap{(}{\sin{\theta}}{)}}^{2}}{)}}^{-1},~\discretionary{}{}{} g^{\theta \theta} = {r}^{-2}}{]}$\end{adjustwidth}

Запишем определение символов Кристоффеля $\Gamma^{\alpha}_{\mu\nu}$:
\begin{python}
Gamma:= \Gamma^{\alpha}_{\mu\nu} = 1/2 g^{\alpha\beta} (\partial_{\nu}{g_{\beta\mu}} + \partial_{\mu}{g_{\beta\nu}} - \partial_{\beta}{g_{\mu\nu}});
\end{python}
\begin{adjustwidth}{1em}{0cm}${}\Gamma^{\alpha}\,_{\mu \nu} = \frac{1}{2}g^{\alpha \beta} \brwrap{(}{\partial_{\nu}{g_{\beta \mu}}+\partial_{\mu}{g_{\beta \nu}}-\partial_{\beta}{g_{\mu \nu}}}{)}$\end{adjustwidth}

Раскроем символы Кристоффеля через метрический тензор. При этом
раскрытие будет применяться только к правой части определения (после
знака равенства). Этот порядок раскрытия определяется опцией
\verb|rhsonly=True|. Функция \verb|evaluate()| имплицитно использует
вызов SymPy для операций с компонентами (это ещё одно преимущество
использования инфраструктуры Python). Для раскрытия тригонометрических
соотношений следует явно обратиться к SymPy: 
\begin{python}
evaluate(Gamma, g, rhsonly=True)
map_sympy(Gamma, "expand_trig");
\end{python}
\begin{adjustwidth}{1em}{0cm}${}\Gamma^{\alpha}\,_{\mu \nu} = \square{}_{\mu}{}_{\nu}{}^{\alpha}\brwrap{\{}{\begin{aligned}\square{}_{\varphi}{}_{\theta}{}^{\varphi}= & {\brwrap{(}{\tan{\theta}}{)}}^{-1}\\[-.5ex]
\square{}_{\theta}{}_{\varphi}{}^{\varphi}= & {\brwrap{(}{\tan{\theta}}{)}}^{-1}\\[-.5ex]
\square{}_{\varphi}{}_{\varphi}{}^{\theta}= & -\sin{\theta} \cos{\theta}\\[-.5ex]
\end{aligned}}{.}$
\end{adjustwidth}

Аналогично для тензора Римана $R^{\alpha}\,_{\beta \mu \nu}$ запишем
его определение и вычислим его компоненты:
\begin{python}
R4:= R^{\rho}_{\sigma\mu\nu} = +\partial_{\mu}{\Gamma^{\rho}_{\sigma\nu}} - \partial_{\nu}{\Gamma^{\rho}_{\sigma\mu}} + \Gamma^{\rho}_{\beta\mu} \Gamma^{\beta}_{\sigma\nu} - \Gamma^{\rho}_{\beta\nu} \Gamma^{\beta}_{\sigma\mu};
substitute(R4, Gamma)
evaluate(R4, g, rhsonly=True);
\end{python}
\begin{adjustwidth}{1em}{0cm}${}R^{\rho}\,_{\sigma \mu \nu} = \partial_{\mu}{\Gamma^{\rho}\,_{\sigma \nu}}-\partial_{\nu}{\Gamma^{\rho}\,_{\sigma \mu}}+\Gamma^{\rho}\,_{\beta \mu} \Gamma^{\beta}\,_{\sigma \nu}-\Gamma^{\rho}\,_{\beta \nu} \Gamma^{\beta}\,_{\sigma \mu}$\end{adjustwidth}
\begin{adjustwidth}{1em}{0cm}${}R^{\rho}\,_{\sigma \mu \nu} = \square{}_{\sigma}{}_{\nu}{}^{\rho}{}_{\mu}\brwrap{\{}{\begin{aligned}\square{}_{\varphi}{}_{\varphi}{}^{\theta}{}_{\theta}= & {\brwrap{(}{\sin{\theta}}{)}}^{2}\\[-.5ex]
\square{}_{\theta}{}_{\varphi}{}^{\varphi}{}_{\theta}= & -1\\[-.5ex]
\square{}_{\varphi}{}_{\theta}{}^{\theta}{}_{\varphi}= & -{\brwrap{(}{\sin{\theta}}{)}}^{2}\\[-.5ex]
\square{}_{\theta}{}_{\theta}{}^{\varphi}{}_{\varphi}= & 1\\[-.5ex]
\end{aligned}}{.}$
\end{adjustwidth}

Тензор Риччи $R_{\alpha\beta}$ вычисляется из тензора Римана:
\begin{python}
R2:= R_{\sigma\nu} = R^{\rho}_{\sigma\rho\nu};
substitute(R2, R4)
evaluate(R2, g, rhsonly=True);
\end{python}
\begin{adjustwidth}{1em}{0cm}${}R_{\sigma \nu} = R^{\rho}\,_{\sigma \rho \nu}$\end{adjustwidth}
\begin{adjustwidth}{1em}{0cm}${}R_{\sigma \nu} = \square{}_{\sigma}{}_{\nu}\brwrap{\{}{\begin{aligned}\square{}_{\varphi}{}_{\varphi}= & {\brwrap{(}{\sin{\theta}}{)}}^{2}\\[-.5ex]
\square{}_{\theta}{}_{\theta}= & 1\\[-.5ex]
\end{aligned}}{.}
$\end{adjustwidth}

И, наконец, вычислим скалярную кривизну $R$:
\begin{python}
R:= R = R_{\sigma\nu} g^{\sigma\nu};
substitute(R, R2)
evaluate(R, g, rhsonly=True);
\end{python}
\begin{adjustwidth}{1em}{0cm}${}R = R_{\sigma \nu} g^{\sigma \nu}$\end{adjustwidth}
\begin{adjustwidth}{1em}{0cm}${}R = 2{r}^{-2}$\end{adjustwidth}

Таким образом, наличие компонентных вычислений для пользователя
выглядит в Cadabra2 как дополнительное свойство \verb|Coordinate| и
несколько функций, основной из которых является функция
\verb|evaluate()|, которая собственно вычисляет значения компонент.

На уровне системы компонентные вычисления реализованы в Cadabra2 через
взаимодействие со скалярной системой компьютерной алгебры, а именно  с
SymPy. 

  \section{Использование графики в Cadabra2}
\label{sec:plot}

Необходимость наличия графики в системе тензорной компьютерной алгебры
вызывает много вопросов. Коротко говоря, по мнению авторов, она просто
не нужна. Однако в данном случае наличие графических возможностей~---
не более, чем дополнительный (в чем-то побочный) эффект реализации
Cadabra2 в рамках экосистемы Python. Поэтому и построение графиков в
Cadabra2 не отличается от построения графиков в Python.

Сначала нужно выбрать библиотеку для построения графиков. В следующем
примере использована популярная библиотека
Matplotlib~\cite{tosi:2009:matplotlib,vaingast:2009:matplotlib,muller:2016:python-ml}. Для
численных расчётов использована библиотека
NumPy~\cite{idris:numpy_cookbook,oliphant:guide_numpy}.

Подключим модули для \verb|matplotlib| и \verb|numpy|:
\begin{python}
import matplotlib.pyplot as plt
import numpy as np
\end{python}

Построим векторное поле, то есть каждой точке пространства (в нашем
случае плоскости) поставим в соответствие вектор с началом в этой
точке. Пусть вектор $f$ имеет вид:
\begin{equation}
  \label{eq:vec_field}
  f^{i} (x,y) = \mqty(\sin x \cos x\\ \cos y).
\end{equation}

Зададим квадратную сетку, на которой и будем вычислять
значения векторного поля. Поскольку отсчёты по осям одинаковы, то будем
использовать только значения для $x$:
\begin{python}
x = np.arange(-2*np.pi, 2*np.pi, 0.1)
u = np.sin(x)*np.cos(x)
v = np.cos(x)
uu, vv = np.meshgrid(u,v)
\end{python}

Функция \verb|meshgrid| из пакета \verb|numpy| создаёт прямоугольную
сетку из двух массивов (в нашем случае из $u$ и $v$).

Теперь, собственно, построим векторное поле с помощью функции
\verb|streamplot| из пакета \verb|matplotlib|:
\begin{python}
fig = plt.figure()
plt.streamplot(x, x, uu, vv, color='black') 
plt.title('Vector field')
plt.xlabel('x')
plt.ylabel('y')
\end{python}

Мы можем как сохранить получившийся график, так и отобразить его в
рабочем блокноте:
\begin{python}
fig.savefig("plot.pdf")
\end{python}
\begin{python}
display(fig)
\end{python}

\begin{figure}
  \includegraphics[width=\linewidth]{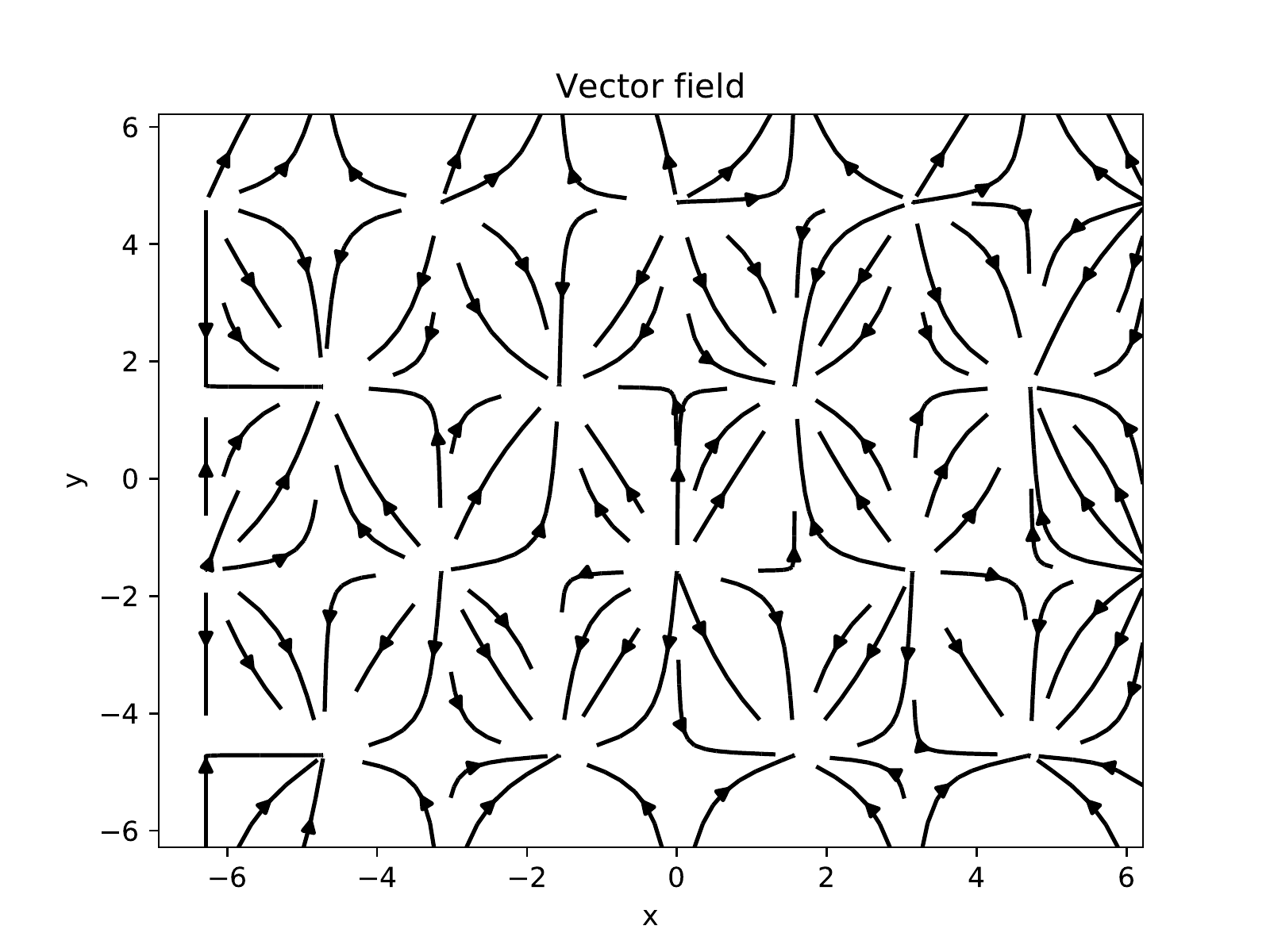}
  \caption{Изображение векторного поля~\eqref{eq:vec_field},
    полученное с помощью matplotlib}
  \label{fig:plot}
\end{figure}

В результате получим изображение векторного поля (см. рис.~\ref{fig:plot}).

Возможно в будущем в Cadabra появятся более полезные приложения, нежели мы
продемонстрировали в этом простом примере.

\section{Заключение}
\label{sec:conclusion}

  В результате можно сделать следующие выводы.  Основным прорывом
  системы Cadabra2 с точки зрения пользователя тензорной системы
  компьютерной алгебры является реализация компонентных тензорных
  вычислений. Таким образом система охватывает весь спектр необходимых
  тензорных операций. С точки же зрения разработчика основным
  нововведением является переписывание системы с использованием языка
  Python и всей экосистемы этого языка. Это позволяет надеяться на
  рост интереса к системе Cadabra2 при решении задач, связанных с
  операциями над тензорами.

\def\acknowledgmentsname{Благодарности}

\begin{acknowledgments}

Публикация подготовлена при поддержке Программы РУДН «5-100»
и при финансовой поддержке РФФИ в рамках научных проектов
\textnumero~16-07-00556, 18-07-00567, 18-51-18005.

\end{acknowledgments}

  \ifthenelse{\boolean{xetex}}
  {\bibliographystyle{ugost2008l}}
  {\bibliographystyle{gost2008l}}

\bibliography{bib/cadabra2/cite}